\documentclass[aps,prl,twocolumn,superscriptaddress]{revtex4}
\usepackage[dvips]{graphicx}
\usepackage{color}
\usepackage{bm}
\usepackage{amssymb}

\usepackage[utf8]{inputenc}

\begin{document}
%\begin{CJK*}{GB}{gbsn}

\title{Dimensionality and irreversibility of field-induced transitions in SrDy$_2$O$_4$}

\author{C.~Bidaud}
\author{O.~Simard}
\affiliation{Département de physique, Université de Sherbrooke, Sherbrooke (Québec) J1K 2R1 Canada}

\author{G. Quirion}
\affiliation{Département de physique, Université de Sherbrooke, Sherbrooke (Québec) J1K 2R1 Canada}
\affiliation{Department of Physics and Physical Oceanography, Memorial University, St. John's, Newfoundland, Canada, A1B 3X7}

\author{B. Prévost}
\author{S. Daneau}
\author{A. D. Bianchi}
\affiliation{Département de Physique, Université de Montréal, Montréal (Québec) H3T 1J4 Canada}

\author{H. A. Dabkowska}
\affiliation{Brockhouse Institute for Materials Research, McMaster University, Hamilton, Ontario L8S 4M1 Canada}

\author{J.~A.~Quilliam}
\email{jeffrey.quilliam@usherbrooke.ca}
\affiliation{Département de physique, Université de Sherbrooke, Sherbrooke (Québec) J1K 2R1 Canada}

\date{\today}

\begin{abstract}
Low temperature ultrasound velocity measurements are presented on the
frustrated spin system SrDy$_2$O$_4$ that allow us to define high
resolution phase diagrams with the magnetic field applied along all
three principal axes. For $H||b$, a region of field-induced long range
order is delimited by a dome of first-order phase transitions. An
unusual magnetization process is observed with significant
irreversibility at very low temperatures when passing between the
low-field spin liquid phase and the long range ordered phase which we
attribute to large energy barriers. For $H||c$, the system appears to
remain effectively one-dimensional, exhibiting two transitions as a
function of magnetic field, but no finite-temperature long range order.

 \end{abstract}
%As expected, the results are essentially featureless for the hard axis, $H||a$.


\maketitle
%\end{CJK*}

An attractive feature of rare-earth magnets is the possibility to grow isostructural materials with many different rare-earth ions, each with distinct total angular momentum, anisotropy and quantum fluctuations, providing a wealth of diverse physical phenomena for a given lattice symmetry. While the pyrochlores are undoubtedly the best studied set of isostructural rare-earth materials~\cite{Gardner2010RMP}, a new group of materials with the formula $AR_2$O$_4$ has recently become a popular field of study~\cite{Petrenko2014}. The $A$ ion can be either Sr or Ba and the rare-earth ion is highly flexible with materials containing Ho~\cite{Fennell2014}, Dy~\cite{Cheffings2013}, Er~\cite{Hayes2011}, and Yb~\cite{QuinteroCastro2012} having recently been studied.  When viewed along the $c$-axis, these materials have a honeycomb structure (see Fig. 1). However, from the point of view of interactions they are likely best represented by a model of two inequivalent sets of frustrated zigzag spin chains with weaker interchain coupling \cite{Petrenko2014}.

One of the most intriguing materials in this class is SrDy$_2$O$_4$ which, in zero-field, shows no indications of long-range order (LRO) down to very low temperatures, suggesting a spin liquid ground state, an intriguing state of matter that supports significant quantum entanglement and exotic excitations~\cite{Balents2010}. In magnetic field, however, thermodynamic measurements have shown evidence of magnetic order~\cite{Hayes2012, Cheffings2013}. In particular, with $H||b$, the magnetization appears to show a plateau from 0.5 to 1.5 T and partial saturation above 2.5 T.  The plateau is found to be roughy 1/3 of the partially saturated moment, suggesting that an up-up-down ($\uparrow\uparrow\downarrow$) state is stabilized by quantum fluctuations~\cite{Hayes2012}. Inelastic neutron scattering measurements and crystal field calculations~\cite{Fennell2014} have allowed for a determination of crystal field levels in SrHo$_2$O$_4$ and SrDy$_2$O$_4$ and indicate that the easy axis of site 1 (the red chains) is the $c$-direction whereas the easy axis of site 2 (the blue chains) is along $b$ (see Fig.~1). Based on bond lengths in the zig-zag chains, they may be well described by a highly frustrated 1d anisotropic next-nearest-neighbor Ising (ANNNI) model~\cite{Selke1988}.

%The fact that the easy axes of adjacent chains are perpendicular presumably plays a major role in mitigating inter-chain interactions and maintaining the 1d character of the material.

\begin{figure}
\begin{center}
\includegraphics[width=3in]{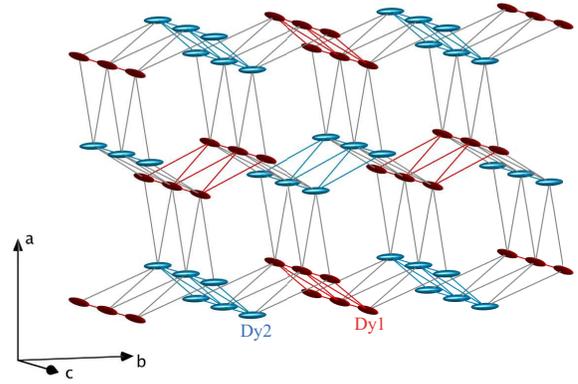}
\caption{Structure of Dy ions in SrDy$_2$O$_4$, with the shape of ellipsoids illustrating the easy axes of the two sites. Site Dy1 (red) has easy its axis along $c$ and site Dy2 (blue) has its easy axis along $b$. The nearest- and next nearest-neighbor bonds form zigzag chains that extend along $c$ while a honeycomb lattice is formed in the $ab$-planes. \label{structure}}
\end{center}
\end{figure}

In this Letter we present a study of SrDy$_2$O$_4$ using ultrasound velocity measurements down to low temperatures, an order of magnitude below what was previously obtained. The sound velocity shows precise anomalies at magnetic phase transitions, providing a high-resolution magnetic phase diagram. Notably, we find that the effective dimensionality of the system (with respect to magnetic degrees of freedom) depends heavily on the direction of the applied magnetic field. We have also carefully studied the effect of field-history, making it possible to identify first-order phase transitions, glassy dynamics and  strong irreversibility when passing between spin liquid and ordered phases.

\begin{figure}
\begin{center}
\includegraphics[width=3.5in]{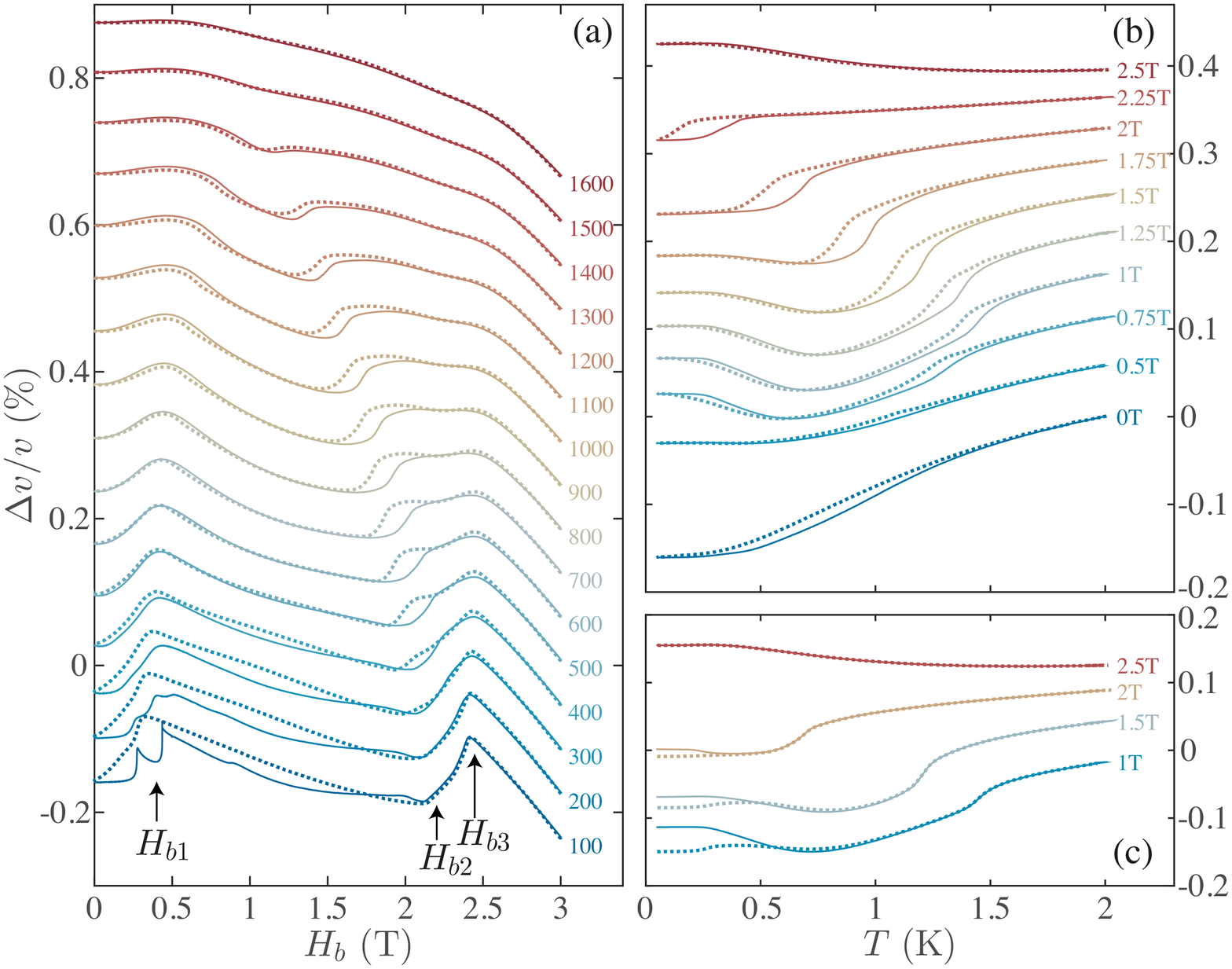}
\caption{(a) Relative change in sound velocity as a function of $H||b$. Dotted lines are for decreasing field, solid lines are for increasing field. Temperatures labels are given in units of mK. (b) Relative change in sound velocity as function of $T$ for various fields $H||b$. Dotted lines are for FC and solid lines are for FCW.  (c) Temperature scans of $\Delta v/v$ for $H||b$ showing the difference between FCW (solid lines) and ZFCW (dotted lines). Curves have been staggered for ease of view.\label{bcurves}}
\end{center}
\end{figure}

A single crystal of SrDy$_2$O$_4$, grown by the floating zone technique, was cut with faces perpendicular to the orthorhombic axes $a$, $b$ and $c$ ($2.26 \times 2.36\times 2.30$ mm$^3$). Transverse piezoelectric transducers were glued on opposite faces of the crystal, oriented such that shear waves were propagated along the $c$-axis (along the chains) with polarization in the $a$-direction. An ultrasonic interferometer was used to measure the velocity of the first elastic transmitted pulse (at 92 MHz) which is proportional to the elastic constant $C_{55}$. The absolute velocity is found by time of flight to be $1440 \pm 15$  m/s.  The sample was cooled to 50 mK in a dilution refrigerator and magnetic fields were applied along the $a$-, $b$- and $c$-axes. Relative changes in velocity, $\Delta v/v$, primarily reflect changes in the restoring forces between atoms due to magneto-elastic coupling which provides a highly sensitive probe of the magnetization and magnetic order parameters~\cite{Quirion2005,Quirion2006}. The $C_{44}$ elastic constant was also measured on a different sample of SrDy$_2$O$_4$ and gave essentially the same results, although with overall larger variations (stronger magnetoelastic coupling).

\begin{figure*}
\begin{center}
\includegraphics[width=6in]{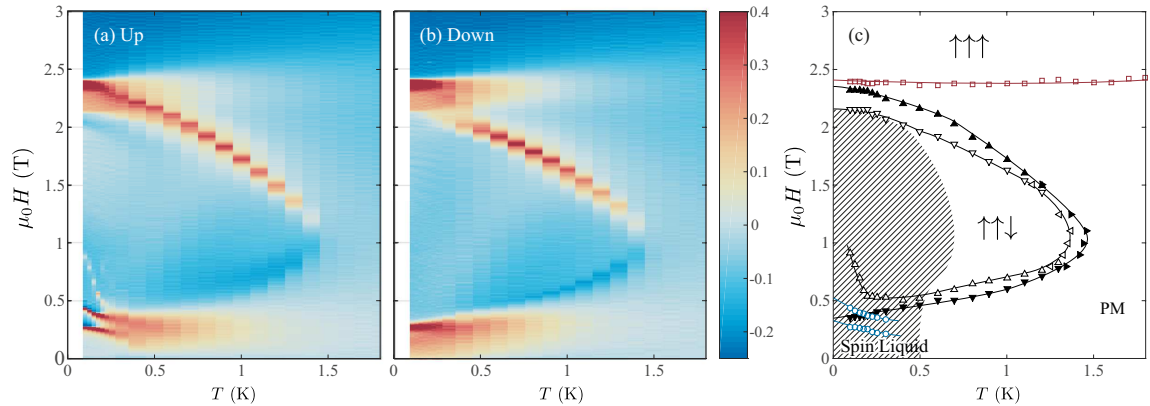}
\caption{Surface plots of the field derivative of sound velocity, $(\partial v/\partial H)/v$ as a function of $H_b$ and temperature for (a) increasing field and (b) decreasing field. In addition to a shift of the position of the dome of LRO (due to hysteresis at 1st order phase transitions), the ``up'' curves demonstrate a great deal of increased structure in the vicinity of $H_{b1}$, with this line of transitions appearing to bifurcate into three separate transitions. (c) Phase diagram for $H||b$. Blue circles are the two initial jumps seen when raising the magnetic field. Black triangles present the first-order phase transitions that delimit what is likely the $\uparrow\uparrow\downarrow$ LRO phase, with the orientation of the triangles indicating the direction of field or temperature sweeps. Red squares illustrate the crossover to full polarization (of Dy2). The hatched region roughly shows where slow (glassy) dynamics are observed.\label{bphasediagram}}
\end{center}
\end{figure*}

$\mathbf{H||b}$ {\bf (blue chains / Dy2)} -- We begin with a discussion of the richest phase diagram, that is for $H||b$ where the field is primarily influencing the spins on the blue chains (Dy2). Anomalies in the sound velocity are primarily associated with changes in the square of the magnetization, $M_{2b}^2$, the development of an  order parameter, $S^2$, and changes in low frequency spin fluctuations. Relative changes in velocity are shown as a function of magnetic field for various temperatures in Fig.~\ref{bcurves}(a). Below $\sim 1.5$~K, we typically observe three separate anomalies where the slope of $v(B)$ changes sign.

The first two critical fields, $H_{b1}$ and $H_{b2}$ are found to converge as the temperature is raised, forming a dome, inside of which is likely a long range ordered (LRO) magnetic state. Clear hysteresis loops indicate that the dome of LRO is delimited by first-order phase transitions. This can be seen in Fig. \ref{bcurves}a where solid (dotted) lines indicate increasing (decreasing) field.  Temperature scans, shown in Fig.~\ref{bcurves}b at various values of $H_b$, are also consistent with a dome of hysteretic, first-order phase transitions. As a function of temperature the lattice is found to soften at the transition (i.e. the velocity is reduced), indicating a negative coupling to the square of the order parameter $S^2$. Hence, in the phase diagram of Fig.~\ref{bphasediagram}c, we have used a minimum in $\partial v/\partial H$ to signal the onset of the order parameter at $H_{b1}$ and the maximum of $\partial v/\partial H$ to signal the disappearance of the order parameter at $H_{b2}$. Given that the magnetization is close to 1/3 of the saturated value in the region under the dome shown in Fig.~\ref{bphasediagram}~\cite{Hayes2012}, this may be an $\uparrow\uparrow\downarrow$ phase on chains formed by Dy2 ions with an easy axis anisotropy pointing along the $b$-axis (see blue chains in Fig.~1). Such a 1/3 magnetization plateau can be seen in simulations of the highly frustrated 1d ANNNI model that is thought to apply to this material~\cite{Yang2014}.

Although we associate anomalies at $H_{b1}$ and $H_{b2}$ with true phase transitions, the origin of the maximum observed in the sound velocity at around $H_{b3} \simeq 2.5$~T seems quite different (Fig.~2a) in that it is quickly smeared out with increasing temperature.  Consequently, we attribute this maximum to a crossover at which the blue chain spins reach their full polarization as shown in magnetization data for H//b \cite{Hayes2012}. Despite full polarization of Dy2 (blue chains) above $H_{b3}$, we nonetheless observe that $\Delta v/v \propto -H_b^2$. In order to account for this significant field dependence of the velocity above $H_{b3}$, we must invoke a negative magnetoelastic coupling with spins on the other chain (red chain / Dy1), which in this regime continue to show a small magnetization that is linear in magnetic field, as measured in Ref.~\cite{Hayes2012}. The leading-order field dependence of the velocity of transverse acoustic waves then simply reduces to $\Delta v/v = k M_{1b}^2 = k \chi_{1b}^2 H^2$~\cite{Quirion2011} where $k$ is a coupling constant and $\chi_{1b}$ is the constant magnetic susceptibility of spins on the red chains (Dy1) with the field along the $b$-axis.

This material's phase diagram shows several regions of hysteresis with a complex field-history dependence. In order to address this complication carefully we have performed temperature scans using three different protocols: 1) cooling in field (FC), 2) warming after field cooling (FCW) and 3) warming after zero-field cooling (ZFCW). The FC and FCW curves primarily show the hysteresis associated with supercooling at the first-order phase transition, as shown in Fig. 2b.  In Fig.~\ref{bcurves}c, we instead observe a strong difference between FCW and ZFCW curves only below $\sim 500$ mK. It seems that cooling the sample in the liquid phase at zero field locks in a significant level of disorder. On increasing the field into the LRO phase, some disorder remains frozen in, probably in the form of domains.

Strangely, below $\sim 250$ mK, the transition at $H_{b2}$ (Fig.~\ref{bcurves}a) no longer shows hysteresis and the same anomalies are seen in both cooling and warming curves. The values of critical fields in this region therefore become somewhat ambiguous. The most peculiar behavior is seen in the field-history dependence of the low-temperature phase transition at $H_{b1}$ which becomes very complex below $\sim 300$ mK. For decreasing magnetic field only one anomaly is observed at 0.36 T (we define the transition as the minimum in $\partial v/\partial H$). However, for increasing magnetic field, $H_{b1}$ appears to bifurcate into three separate anomalies which, at 100 mK, are found at 0.27, 0.44 and 0.91 T. This is evident in the surface plot of Fig.~\ref{bphasediagram}a, where we show $(\partial v/\partial H)/v$, as well as in the phase diagram presented in Fig.~\ref{bphasediagram}c.

This unusual irreversibility can likely be explained by the slow dynamics that are observed at low-field, for example as a FC-ZFC splitting in magnetization measurements~\cite{Hayes2012}. We also find a difference between cooling and warming curves which indicate slow dynamics in the liquid phase, but this difference depends on the sweep rate and does not show a pronounced dip in velocity at a specific temperature as would be expected for a spin glass transition~\cite{Huang1982}. Likely, this system can best be described as a ``classical'' spin liquid, with slow dynamics and minimal quantum fluctuations.

Similar to spin ice, SrDy$_2$O$_4$ in zero-field may progressively freeze into one of many possible degenerate disordered ground states with relatively short-range correlations. Many rare-earth Ising systems like this one can even manifest extremely slow spin dynamics in paramagnetic~\cite{Quilliam2012hoylf, Schechter2008} and spin liquid~\cite{Snyder2001,Quilliam2011HTO} phases due to very small quantum tunnelling matrix elements, although this is not universally true (see for instance Ref.~\cite{Rau2015}) . When the magnetic field is increased, one selects a particular ground state (likely an $\uparrow\uparrow\downarrow$ state), but the transition from the disordered liquid state to LRO is a violent one as significant energy barriers must be overcome in order to reconfigure small domains into extended order. This transition is evidently achieved in three distinct steps. Only once the field reaches close to 1 T, much higher than the 0.5 T transition that is measured at slightly higher temperature, is the system finally able to achieve LRO. Even then, a domain structure must persist as evidenced by the significant difference between ZFCW and FCW curves in Fig. 1c. On the other hand, when decreasing the magnetic field across $H_{b1}$, the transition is able to proceed much more smoothly as the $\uparrow\uparrow\downarrow$ configuration may be much closer to one of the many possible, short-range ordered ground states permissible in the spin liquid phase and in this case a single anomaly is observed. Fennell \emph{et al.}~\cite{Fennell2014} have demonstrated 1-dimensional, short-range, $\uparrow\uparrow\downarrow\downarrow$ correlations in this phase, and it would be useful to explore theoretically how the system passes between such a short-range ordered state and $\uparrow\uparrow\downarrow$ order. 

Strong irreversibility in field-sweeps has also been observed in spin ice, the most famous example of a classical spin liquid.  Specifically, with the field along the $[111]$ direction, a strong irreversibility was seen well below any phase transitions with magnetization~\cite{Slobinsky2010}, ultrasound velocity~\cite{Erfanifam2011}, and thermal conductivity~\cite{Fan2013} measurements. Again, significant energy barriers, in this case connected to hopping magnetic monopoles~\cite{Castelnovo2008, Jaubert2009}, make it difficult to increase the level of order in the system, leading to avalanches and irreversibility.

\begin{figure}
\begin{center}
\includegraphics[width=3.5in]{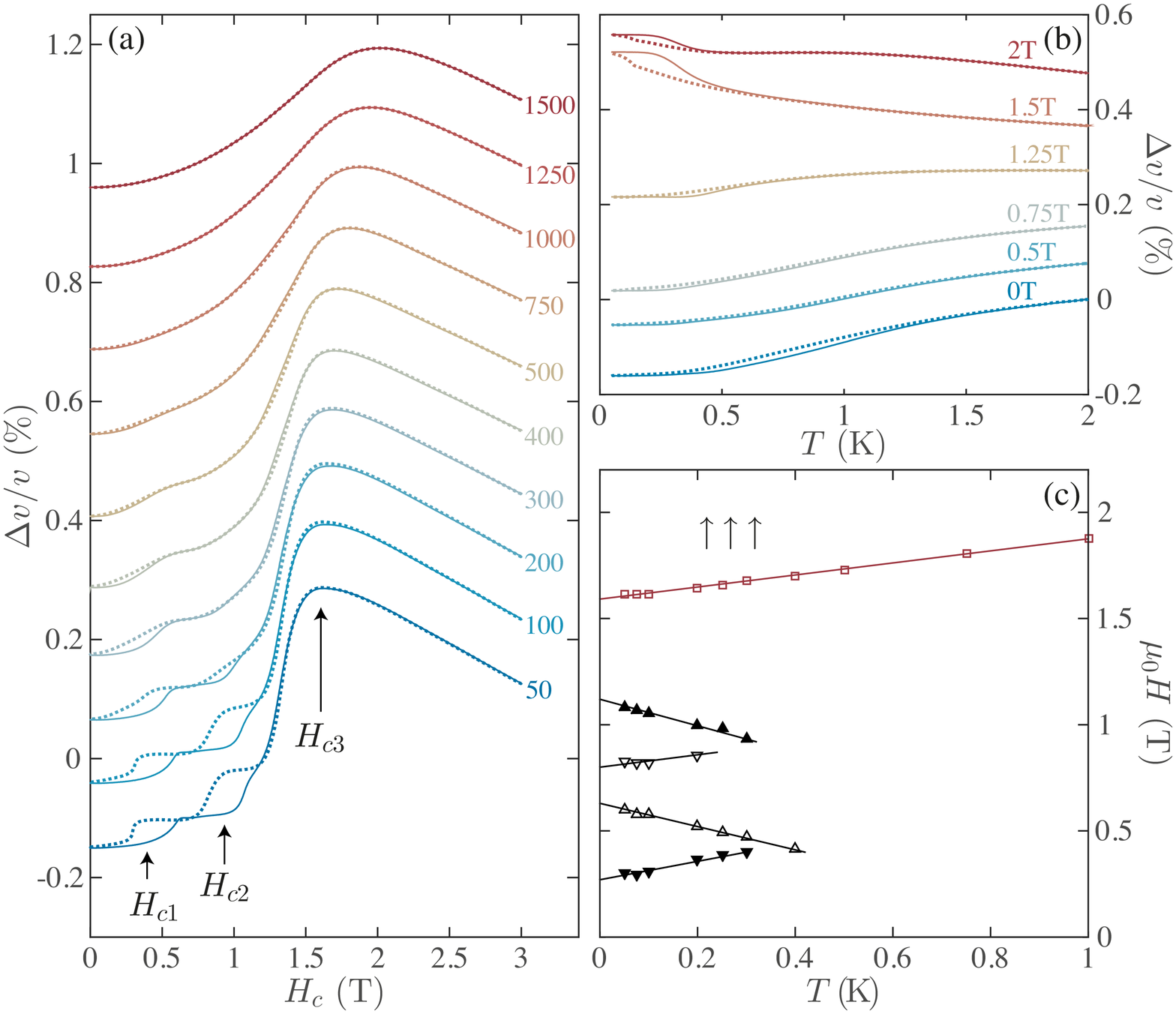}
\caption{Relative change in sound velocity for $H||c$ (a) as a function field and (b) as a function of temperature. Solid lines correspond to increasing field or temperature and dotted lines correspond to decreasing field or temperature. Curves have been staggered for ease of view. (c) Phase diagram for $H||c$. The orientation of the triangles indicate direction of field sweeps. The red squares show a crossover to a ferromagnetic (fully polarized) phase.\label{cfigure}}
\end{center}
\end{figure}

$\mathbf{H||c}$ {\bf (red chains / Dy1)} -- Relative sound velocity is shown as a function of $H||c$ in Fig.~\ref{cfigure}(a).  These field sweeps show qualitatively similar behavior to those with $H||b$ including two hysteretic anomalies and a maximum in the velocity.  Quantitatively speaking, these transitions occur at lower field values: $H_{c1} = 0.295$ (0.595) T and $H_{c2} = 0.821$ (1.07) T for decreasing (increasing) field at 50 mK. 
Saturation of the red chain magnetization gives a very pronounced maximum at $H_{c3} \simeq 1.3 T$, as opposed to $H_{b3} \simeq 2.4$ T. Above this point we again have a negative $H^2$ background. In this case, we speculate that the background comes from the uniform magnetization of Dy2, $M_{2c}$, which should be roughly linear for this field direction.

With increasing temperature, the $H_{c1}$ and $H_{c2}$ anomalies disappear much more quickly than their counterparts for $H||b$.  The striking difference in the $H||c$ data is the lack of phase transitions as a function of temperature. In other words, there does not appear to be a dome of long range order for $H||c$. In particular, this can be seen in the temperature sweep at 0.75 T in Fig.~\ref{cfigure}b (in between the critical fields) which is quite featureless. Thus it appears that the Dy1 sites (red chains) never develop true LRO at a finite temperature, although at the lowest temperatures we may be approaching a zero-temperature LRO that would be consistent with a strongly anisotropic 1d system. It appears that due to some details of the system's Hamiltonian  and crystal field levels, the effective dimensionality of this material varies with field direction. Clear 3-dimensional LRO is found when $H||b$, but the system remains effectively 1-dimensional  when $H||c$.

It is important to highlight the fact that significant slow (glassy) dynamics persist below 500 mK for essentially all values of magnetic field when $H||c$. This glassiness becomes particularly evident above 1.5 T, as can be seen in the hysteresis loops in Fig.~\ref{cfigure}, but this progressive spin freezing does not develop at a sharp glass transition. This is a point that deserves further research with techniques like ac susceptibility or muon spin rotation ($\mu$SR) which are better suited to studying slow spin dynamics.

%Even though $H||c$, the glassiness likely comes primarily from the blue chains which are not heavily affected by the magnetic field and whose spin dynamics continue to slow below 500 mK.

$\mathbf{H||a}$ -- When the magnetic field is applied in the $a$-direction, the obtained curves are relatively featureless, as shown in the supplemental information document~\footnote{Supplementary Material}, apart from some weak hysteresis at the lowest temperatures. $\Delta v/v$ initially increases as $H_a^2$, indicating a quadratic coupling to a linear $M_a(H_a)$. This is not surprising given that the $a$-direction is a hard axis for both types of spin. Some levelling off (beginning of saturation) is noticed by 3 T, as is the case in the magnetization data~\cite{Hayes2012}.

There is a certain degree of similarity between this material and others in the SrLn$_2$O$_4$ family in that one often encounters very different physics on the two different rare-earth sites (the red or the blue chains)~\cite{Young2013}. With the exception of SrYb$_2$O$_4$~\cite{QuinteroCastro2012}, LRO occurs on at most one of the rare-earth sites in all members of this family of materials ~\cite{Aczel2015}. In SrHo$_2$O$_4$, for instance, the red chains appear to develop fairly conventional long range N{\'{e}}el order ($\uparrow\downarrow\uparrow\downarrow$) whereas the blue chains develop only short range order with a double-N{\'{e}}el ($\uparrow\uparrow\downarrow\downarrow$) configuration~\cite{Wen2015}. In both SrHo$_2$O$_4$ and SrDy$_2$O$_4$, the crystal field energy gap is larger for site 2 (blue chains) than for site 1 (red chains)~\cite{Fennell2014} and Wen \emph{et al.}~\cite{Wen2015} argue that this leads to slower dynamics in the blue chains. It is proposed that this slow dynamics, as well as the topological nature of domain walls in the double-N{\'{e}}el state, leads to the blue chains being easily quenched into a state of short range order (SRO). The red chains have much faster dynamics (especially in the Ho compound) and therefore do not so easily loose ergodicity and do find a state of LRO~\cite{Wen2015}. This same logic evidently does not apply to SrDy$_2$O$_4$, however, which has a rather similar crystal field scheme, but only seems to show LRO on the blue chains under applied field $H||b$. A similar coexistence of LRO and SRO has been observed in SrEr$_2$O$_4$~\cite{Hayes2011}, BaNd$_2$O$_4$ shows LRO of the double-N{\'{e}}el ($\uparrow\uparrow\downarrow\downarrow$) variety on only one of the chains~\cite{Aczel2014} and SrTb$_2$O$_4$ has incommensurate order on only one of the chains~\cite{Li2014SrTb2O4}.Two other systems in the family, BaTb$_2$O$_4$~\cite{Aczel2015} and SrTm$_2$O$_4$~\cite{Li2014}, do not show any magnetic order in zero-field but their field-induced phase diagrams have not yet been explored.

In conclusion, this work has elaborated the first detailed phase diagrams of SrDy$_2$O$_4$ for magnetic field along the $b$- and $c$-directions and for temperatures down to 50 mK (Fig.~\ref{bphasediagram}c and Fig.~\ref{cfigure}c). Our results are consistent with those of magnetization~\cite{Hayes2012} and specific heat~\cite{Cheffings2013} measurements, but extend to lower temperatures and provide much higher resolution. Whereas a clear dome of long range order is found for $H||b$ for spins with a $b$-axis anisotropy (blue chains), the red chains appear to remain disordered for all magnetic fields, with LRO only possibly occurring in the zero-temperature limit when $H||c$. This suggests a difference in effective dimensionality depending on the direction in which the magnetic field is applied. A very strong irreversibility at the first critical field value seems to be consistent with a transition into or out of a disordered spin liquid with large energy barriers impeding ordering of the spins as the field is increased. 

Finally, using the highly sensitive ultrasound velocity technique, we have shown that there is no ordering anomaly in zero-field down to 50 mK, therefore providing strong evidence of frustration and a likely spin liquid ground state. Previous thermodynamic measurements have only reached $\sim 500$ mK, which is just slightly lower than the ordering transition in SrHo$_2$O$_4$ at $\sim 600$ mK, for instance~\cite{Hayes2012,Cheffings2013}. Slow dynamics suggest a ``classical'' spin liquid with only short range correlations~\cite{Fennell2014} along with large energy barriers and a progressive slowing of spin dynamics, similar to spin ice for example~\cite{Snyder2001}. This material's rich behavior provides an interesting challenge for theory. It remains to be seen whether the origin of the spin liquid phase and the unusual transitions to LRO can be understood within the context of fairly simple 1d ANNNI models or whether additional complications like the long-range dipolar interaction, interchain couplings and additional crystal field levels are essential to modeling the magnetism in SrDy$_2$O$_4$.

\begin{acknowledgements}
We are grateful to M. Castonguay for extensive technical support and we acknowledge valuable discussions with M. Kenzelmann, M. Gingras and A. Aczel. This work was supported by NSERC and the RQMP which is funded by the FRQNT. 
\end{acknowledgements}

%\bibliographystyle{h-physrev3}
%\bibliography{SDOrefs.bib}

\end{document}